\documentstyle[12pt,aasms4]{article}

\slugcomment{\apj in press}

\lefthead{Xiangdong Shi \& Michael S. Turner}
\righthead{Expectations for the Difference Between Local and
Global Measurements of the Hubble Constant}

\begin{document}
\title{Expectations for the Difference Between Local and Global Measurements
of the Hubble Constant}

\author{Xiangdong Shi}
\affil{shi@doheny.ucsd.edu\\
Department of Physics, University of California, La Jolla, 
CA 92093-0319}

\and

\author{Michael S. Turner}
\affil{mturner@oddjob.uchicago.edu \\
Departments of Astronomy \& Astrophysics and of Physics\\
Enrico Fermi Institute, The University of Chicago\\
Chicago, IL 60637-1433\\
NASA/Fermilab Astrophysics Center\\
Fermi National Accelerator Laboratory\\
Batavia, IL~~60510-0500}

\begin{abstract}
There are irreducible differences between the Hubble constant
measured locally and the global value.  They are due to density perturbations
and finite sample volume (cosmic variance) and finite number of objects
in the sample (sampling variance).  We quantify these differences
for a suite of COBE-normalized CDM models that are consistent
with the observed large-scale structure.  For small samples of
objects that only extend out to 10,000 km/sec, the variance
can approach 4\%.  For the largest samples
of Type Ia supernovae (SNeIa), which include about 40
objects and extend out to almost 40,000 km/sec,
the variance is $1 -2 \%$ and is
dominated by sampling variance.  Sampling and cosmic variance may
be an important consideration in comparing local determinations
of the Hubble constant with precision determinations
of the global value that will be made from high-resolution maps
of CBR anisotropy.

\end{abstract}


\section{Introduction}

On the largest scales ($\gg 100\,$Mpc) the Universe is well described by
homogeneous and isotropic expansion satisfying the Friedmann equation.
The global rate of the expansion at the current epoch is defined to be
the Hubble constant ($\equiv H_0$), the fundamental parameter of cosmology
that sets the size and age of the Universe.  Many methods have been
used to measure $H_0$ (see, e.g., Freedman 1996).  Currently,
the use of SNeIa as standard candles yields the smallest
estimated measurement error (Riess, Press and Kirshner 1996;
Hamuy et al 1996; Saha et al 1996).

On small scales ($\la 100$ Mpc), the Universe is significantly
inhomogeneous.  Because density fluctuations give rise to deviations
from isotropic and homogeneous expansion (peculiar velocities)
the expansion cannot be characterized by a universal expansion
rate and measurements within a small, finite
region will yield a local expansion rate ($\equiv H_{\rm L}$)
which is not identical to the global expansion rate.  The difference
arises from two factors:  the finite physical size of the sample
(Turner, Cen \& Ostriker 1992; Nakamura and Suto 1995;
Shi, Widrow and Dursi 1996; Turner 1997) and
the limited number of objects in the sample.  Because of peculiar
velocities, the average expansion rate in a finite volume
is different from the global expansion (cosmic variance).
Moreover, because only a small number of points within the
volume are sampled the expansion rate defined by these points can
deviate from the average expansion rate for the volume (sampling variance).
These effects are different
from measurement error, which can be reduced by
better measurements and/or better standard candles.
Cosmic and sampling variance can only be reduced by increasing the
sample volume and the sampling density.

In this {\it Letter}, we quantify cosmic and sampling variance
for two samples of SNeIa and a cluster sample with
Tully-Fisher distances.  In so doing we use a suite of
COBE-normalized CDM models that are consistent with the measurements
of the level of inhomogeneity on scales less than about 300\,Mpc
and which should therefore provide a reasonable estimate for the
variance that arises due to inhomogeneity in the Universe.

\section{Methodology}
The deviation of the local expansion rate measured by an observer at
position ${\bf r}$ from the global Hubble constant is given by
\begin{equation}
{H_L({\bf r})-H_0\over H_0}=
{\delta H({\bf r})\over H_0}=\int
{{\bf v}({\bf r}^\prime-{\bf r})\cdot ({\bf \hat r}^\prime-{\bf\hat r})
\over H_0|{\bf r}-{\bf r}^\prime|}W({\bf r}^\prime-{\bf r})\, d^3r^\prime ,
\label{eq1}
\end{equation}
where ${\bf v}({\bf r}^\prime-{\bf r})\cdot ({\bf\hat r}^\prime-{\bf\hat r})$
is the measured radial peculiar velocity at ${\bf r}^\prime$ and
$W({\bf r}^\prime-{\bf r})$ is the window function that characterizes
the sample volume and sampled points within the volume (more below).
The quantity ${\bf v}({\bf r}^\prime-{\bf r})\cdot
({\bf\hat r}^\prime-{\bf\hat r})$ consists
of two parts, the actual radial peculiar velocity
and its measurement error, which is roughly the measurement error
of the distance scaled by $H_0$.  
The measurement error can be shrunk by improving
distance measurements, but the real radial peculiar velocity is an intrinsic
deviation from the Hubble flow that is determined
by the underlying density fluctuations.  It is its contribution
to ${\delta H({\bf r})/H_0}$ that is irreducible.

The peculiar-velocity field depends upon the underlying power spectrum
of density perturbations, which in turn, depends upon the cosmological
scenario.  We shall investigate a number of CDM models.
In linear-perturbation theory,
\begin{equation}
{\bf v}({\bf r})~=~{H_0\Omega_M^{0.6}\over 2\pi}
\int\,{{\bf r}-{\bf r}^\prime\over |{\bf r}-{\bf r}^\prime|^3}
\,\delta({\bf r}^\prime)\,d^3r^\prime
\end{equation}
where $\delta({\bf r}^\prime)$ is the density fluctuation
$[\rho({\bf r}^\prime)-\bar\rho]/\bar\rho$ and $\Omega_M$ is the
fraction of critical density in matter that clusters.
Taking the Fourier transform we find,
\begin{equation}
{\delta H({\bf r})\over H_0}~=~\Omega_M^{0.6}\int{d^3k\over (2\pi)^{3/2}}
\,\delta({\bf k})\,{{\bf k}\cdot {\bf Z({\bf k})}\over k^2}e^{i{\bf k\cdot r}},
\end{equation}
where $\delta({\bf k})\equiv \left (2\pi\right )^{-3/2}
\int d^3r\delta ({\bf r})e^{i{\bf k\cdot r}}$ and
${\bf Z}({\bf k})\equiv \int d^3r W({\bf r}){\bf\hat r}/r
\,e^{i{\bf k\cdot r}}$.
The variance of ${\delta H({\bf r})/H_0}$ is
\begin{equation}
\Bigl\langle\Bigl({\delta H\over H_0}\Bigr)^2\Bigr\rangle^{1/2}
=\Omega_M^{1.2}\int {d^3k\over (2\pi)^3}\,
{P({\bf k})\over k^2}\,\vert {\bf Z}({\bf k})\cdot{\bf\hat k}\vert^2.
\end{equation}

For a top-hat, spherical window function,
\begin{equation}
{\bf Z}({\bf k})=3\,\frac{ \sin(kR) - 
\int_0^{kR} dx \sin(x)/x}{(kR)^3}\,{\bf k} \, ,
\label{tophat}
\end{equation}
where $W({\bf r}^\prime-{\bf r})=\Theta(R-|{\bf r}^{\prime}-{\bf r}|)/
(4\pi R^3/3)$, $\Theta (x)$ is the step function and
$R$ is the radius of the top-hat sphere (Shi, Widrow and Dursi 1996).
Since the top-hat spherical window function samples every point
within the sphere, $\langle({\delta H({\bf r})/H_0})^2
\rangle^{1/2}$ reflects only the cosmic variance associated
with the finite volume of the spherical sample.

Real data sets do not sample every point in space, rather they sample
a number of objects with positions ${\bf r}_q$ and redshifts $z_q$.  The
radial peculiar velocities of these objects are
\begin{equation}
{\bf v}_q\cdot{{\bf\hat r}}_q=cz_q - H_0 r_q\,
\end{equation}
with uncertainties $\sigma _q$ which are essentially the uncertainties in
distances $r_q$ scaled by $H_0$ because the uncertainties in $z_q$ are
relatively small.  Random motions due to local nonlinearities, characterized
by a one-dimensional standard velocity dispersion $\sigma_*$,
may be added to $\sigma_q$ in quadrature.  Typically $\sigma_*\sim 10^2$
km/sec, and is therefore negligible when the sample depth is as large as
$\sim 10^4$ km/sec.  Corrections to the linear Hubble law due to
deceleration can also be safely ignored for samples with $z\ll 1$.
With these approximations it follows that (Shi 1997)
\begin{equation}
Z^i({\bf k})=
   B^{-1}\left[\sum_q {r_q^i\over\sigma_q^2}e^{i{\bf k}\cdot{\bf r}_q}-
   (A-RB^{-1})^{-1}_{jl}\left( \sum_q{\hat r_q^i\hat r_q^j\over\sigma_q^2}
   e^{i{\bf k}\cdot{\bf r}_q}-
   B^{-1}\sum_q\sum_{q^\prime}{r_q^ir_{q^\prime}^j\over
   \sigma_q^2\sigma_{q^\prime}^2}e^{i{\bf k}\cdot{\bf r}_q}\right)
   \sum_{q^{\prime\prime}}{r_{q^{\prime\prime}}^l
   \over\sigma_{q^{\prime\prime}}^2}\right] ,
\label{window}
\end{equation}
where
\begin{equation}
A_{ij}=\sum_q{\hat r_q^i\hat r_q^j\over\sigma_q^2},\quad
R_{ij}=\sum_q\sum_{q^\prime}{r_q^ir_{q^\prime}^j
\over\sigma_q^2\sigma_{q^\prime}^2},\quad
B=\sum_q{r_q^2\over\sigma_q^2} ,
\label{eq8}
\end{equation}
indices $q$, $q^\prime$ and $q^{\prime\prime}$ denote summation over objects,
and indices $i$, $j$, $l$, $m$ are spatial indices that run from 1 to 3.
Now $\langle({\delta H({\bf r})/H_0})^2\rangle^{1/2}$ includes
both cosmic variance and sampling variance.

\section{Results}
Using Eqs.~(\ref{eq1}) to (\ref{tophat}) we have calculated
the cosmic variance portion of
$\langle({\delta H({\bf r})/H_0})^2\rangle^{1/2}$ with
$R=$7,000, 10,000, 15,000, 20,000, 25,000, and 30,000 km/sec (see Fig.~1).
The underlying cosmological models are a suite of COBE-normalized
CDM models that are consistent with large-scale structure on
scales from about 300\,Mpc to 0.1\,Mpc (Dodelson, Gates, and Turner
1996).  The CDM models include a model with a low value of the
Hubble constant, with significant tilt, with 20\% light neutrinos,
with additional radiation, and with a cosmological constant.
In addition, we have included an open CDM model that is consistent
with large-scale structure measurements and for completeness,
a standard CDM model, which has excessive inhomogeneity on scales
less than 300\,Mpc.  These models should serve well to span theoretical
expectations for the level of inhomogeneity on the scales relevant
for local variations in the Hubble constant.  Their cosmic parameters are
summarized in Table 1.

The cosmic variance of $H_L$ at $R=7,000$ km/sec is
significant, ranging from about 2\% to almost 4\%.  At $R=10,000$ km/sec
it has fallen to 1$\%$ to 2$\%$, and quickly declines to below $1\%$ at
$R=15,000$ km/sec for all models.  At a depth of $R=30,000$ km/sec,
which is reached by SNeIa, the cosmic variance is only about $0.2\%$.

Using Eq.~(\ref{eq8}) we have calculated
$\langle({\delta H({\bf r})/H_0})^2\rangle^{1/2}$
for volumes that are sampled by a finite number of points,
so that both cosmic and sampling variance
are included.  For definiteness we use the SNeIa sample of
Riess et al (1997), for which $H_L=65$ km/sec/Mpc, the SNeIa sample of
Hamuy et al (1996), for which $H_L=63.1\pm 3.4\pm 2.9$ km/sec/Mpc, and
the Tully-Fisher sample of 36 clusters used for the template
Tully-Fisher relation in the Mark III catalogue (Willick et al 1997).
Our results are compiled in Table 2.

The intrinsic variance of $H_L$ measured in
the two SNeIa samples is around $1\%$, far less than measurement
error.  Although the SNeIa sample of Riess et al (1997) is deeper and has
more objects than that of Hamuy et al (1996), its effective depth
is not as large because many of the SNeIa are nearby.
Due to its shallow depth, the Tully-Fisher cluster sample
has a larger intrinsic variance, between 1.5\% and 3\%.

Figure 2 illustrates the effect of finite sampling for
the SNeIa sample of Riess et al (1997).  Comparing this plot to
Fig.~1, it can be seen that the cosmic + sampling
variance is more than twice the cosmic variance.  At present,
sampling variance dominates the intrinsic variance
for the SNeIa samples.  Since sampling variance scales
roughly as the inverse square root of the number of objects,
it can be shrunk to less $1\%$ for all viable CDM models
if the number of SNeIa is doubled.

\section{Summary}

There are intrinsic and irreducible differences between the locally measured
value of the Hubble constant and the global value.  They arise
due to finite sample volume (cosmic variance) and finite
sample size (sampling variance) and can of course be of either sign.
Cosmic variance and sampling variance cannot be reduced by better
measurements or better standard candles.

We have calculated the theoretical expectations for a suite of
COBE-normalized CDM models that are consistent with measurements
of large-scale structure on the scales that give rise to
the cosmic variance portion.  For samples that only extend out to
7,000 km/sec the cosmic variance alone can be close to 4\%; for
samples of around 30 objects that extend out to 10,000 km/sec
cosmic + sampling variance is between 2\% and 4\%.  For samples
of around 40 objects that extend out to 40,000 km/sec the
the total variance is between 0.5\% and 1.5\%, with the cosmic
variance contribution being less than 0.25\%.

As local measurements of the Hubble constant become more precise,
cosmic variance and sampling variance will become a larger
portion of the error budget and may be important when comparing
local measurements with the better than 1\% determinations of
the Hubble constant anticipated from high-resolution maps of
CBR anisotropy (Jungman et al 1996).

The authors thank Adam Riess for providing the positions of their
unpublished SNeIa.  X.S. is supported by grants NASA NAG5-3062
and NSF PHY95-03384 at UCSD. M.S.T. is supported by DoE (at Chicago
and Fermilab) and by the NASA through grant NAG 5-2788 at Fermilab.

\begin{deluxetable}{ccccccccc}
\footnotesize
\tablecaption{Parameters of seven CDM models\tablenotemark{*}}
\tablewidth{0pt}
\tablehead{\colhead{} && \colhead{sCDM} & \colhead{$h=0.4$ CDM}
& \colhead{tCDM} & \colhead{$\nu$CDM} & \colhead{$\tau$CDM}  
& \colhead{$\Lambda$CDM} & \colhead{oCDM}}
\startdata
$\Omega_{\rm TOT}$& &1   &1   &1   &1   &1   &1   &0.4   \nl
$\Omega_M$        & &1   &1   &1   &1   &1   &0.4 &0.4   \nl
$\Omega_B$        & &0.1 &0.16&0.08&0.07&0.08&0.06&0.07  \nl
$\Omega_\nu$      & &0   &0   &0   &0.2 &0   &0   &0     \nl
$h$               & &0.5 &0.4 &0.55&0.6 &0.55&0.65&0.6   \nl
$n$               & &1   &1   &0.7 &1   &0.95&1   &1.1   \nl
\enddata
\tablenotetext{*}{Models and are from Dodelson, Gates and Turner (1996).
Their power spectra are based upon Bardeen et al
(1986) (transfer function), Bunn and White (1997) (COBE normalization),
Sugiyama (1995) (effect of $\Omega_B$ on transfer function),
Ma (1996) ($\nu$CDM transfer function).  The oCDM model is
White and Silk (1996) (oCDM).  With the exception of standard
CDM, which is included only for completeness, all models are
consistent with measures of large-scale structure on scales
from 300\,Mpc to 0.1\,Mpc.}
\end{deluxetable}

\begin{deluxetable}{cccc}
\footnotesize
\tablecaption{Cosmic + sampling variance for three samples.}
\tablewidth{500pt}
\tablehead{\colhead{} & \colhead{SNeIa}
& \colhead{SNeIa}
& \colhead{Tully-Fisher} \nl
\colhead{Reference}& \colhead{Riess et al 1997}
& \colhead{Hamuy et al 1996}
& \colhead{Willick et al 1997} \nl
\colhead{Maximal depth} & \colhead{37,000 km/sec} & \colhead{30,000 km/sec}
& \colhead{11,000 km/sec}   \nl
\colhead{$\pi/k_{\rm peak}$\tablenotemark{*}}
& \colhead{$\sim$21,000 km/sec} & \colhead{$\sim 27,000$ km/sec}
& \colhead{$\sim 10,000$ km/sec}  \nl
\colhead{\# of objects} & \colhead{44} & \colhead{26} & \colhead{36}}
\startdata
sCDM       & 1.4$\%$ & 1.4$\%$ & 3.1$\%$  \nl
$h=0.4$ CDM& 1.1$\%$ & 1.1$\%$ & 2.6$\%$  \nl
$\tau$CDM  & 1.0$\%$ & 0.9$\%$ & 2.3$\%$  \nl
$\nu$CDM   & 1.3$\%$ & 1.3$\%$ & 3.1$\%$  \nl
tCDM       & 0.8$\%$ & 0.8$\%$ & 1.8$\%$  \nl
$\Lambda$CDM & 1.0$\%$ & 0.9$\%$ & 2.4$\%$\nl
oCDM       & 0.7$\%$ & 0.6$\%$ & 1.6$\%$  \nl
\enddata
\tablenotetext{*}{$k_{\rm peak}$ is the wave number where
$\vert {\bf Z}({\bf k})\cdot{\bf\hat k}\vert^2$ reaches maximum.}
\end{deluxetable}

\newpage
\noindent Figure Captions

\noindent Fig.~1:  Cosmic variation as a function of the radius of the
sample volume for the seven COBE-normalized CDM models considered.

\noindent Fig.~2:  Cosmic + sampling variance for the SNeIa sample of Riess
et al (1997).  The numbers indicate the number of objects within
the spherical sample volume.

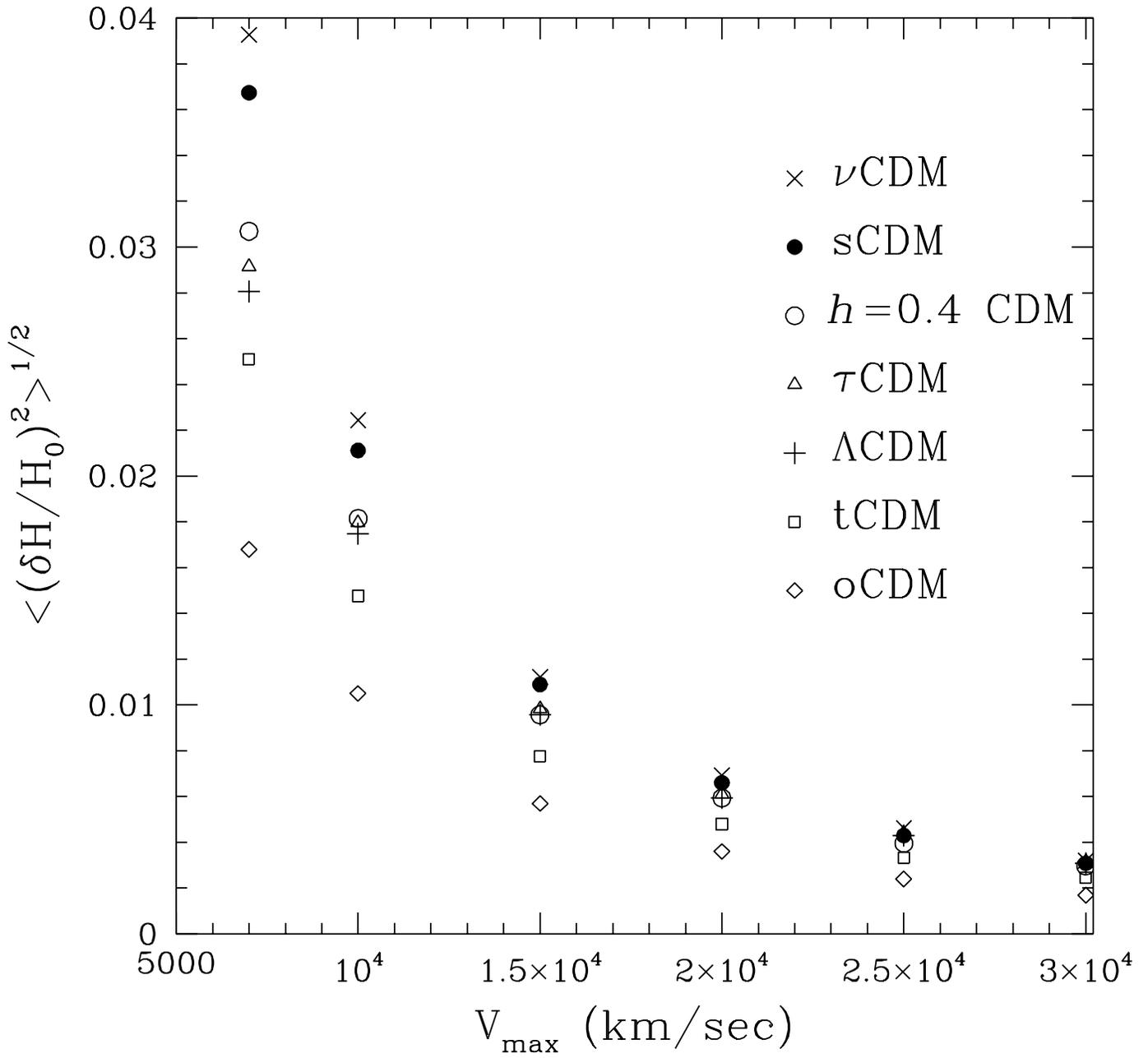
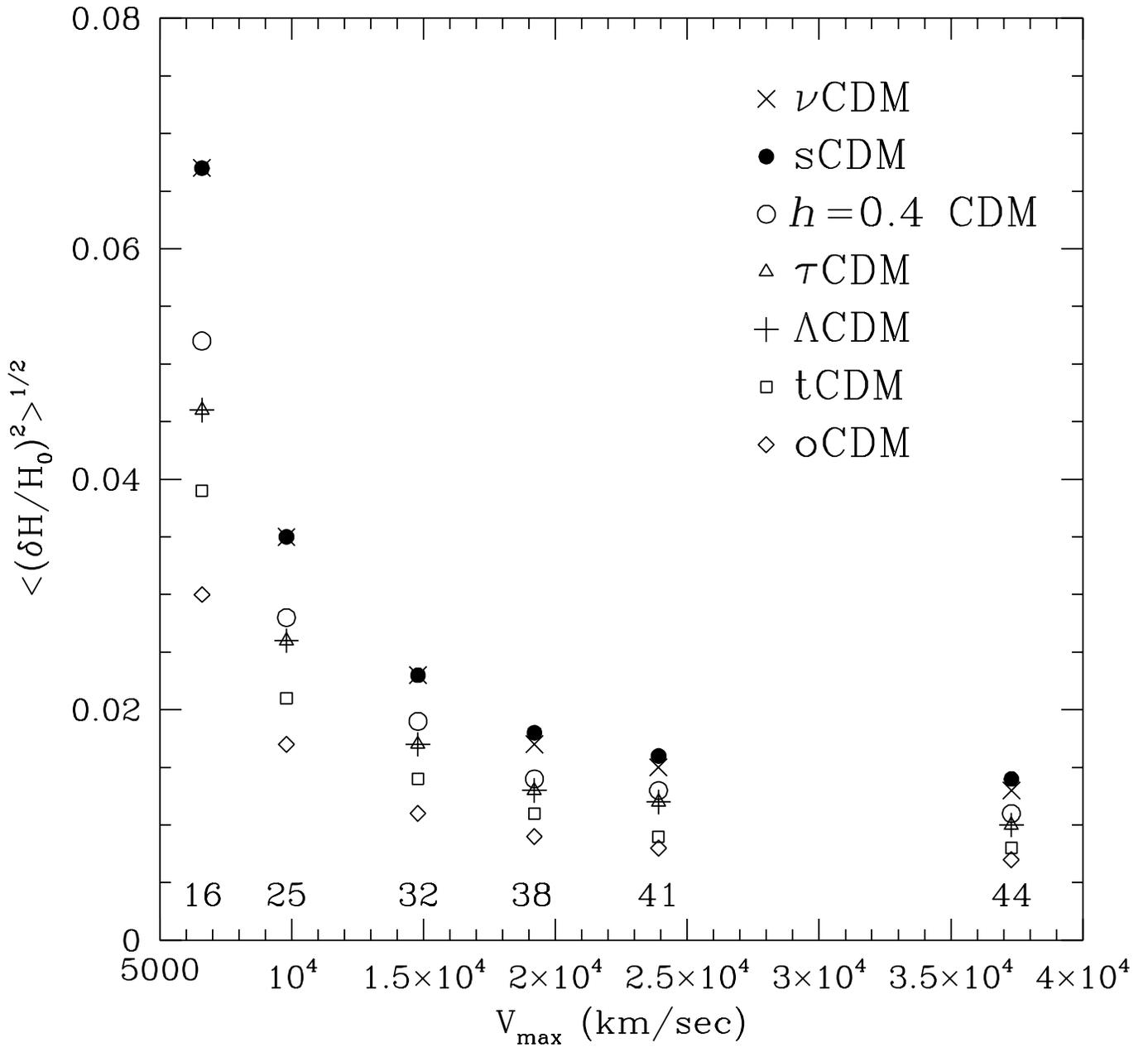
\end{document}